\documentclass[showpacs,preprintnumbers,
amsmath,amssymb,aps,prd]{revtex4}
\usepackage{amsfonts}
\usepackage{amsfonts,graphics,epsfig,subfigure}

\begin{document}

\title{Ehrenfest scheme for $P-V$ criticality of higher dimensional charged black holes, rotating black holes and Gauss-Bonnet AdS black holes}
\author{Jie-Xiong Mo $^{1,2}$ \footnote{mojiexiong@gmail.com}, Wen-Biao Liu $^1$ \footnote{corresponding
author: wbliu@bnu.edu.cn}}

 \affiliation{$^1$ Department of Physics, Institute of Theoretical Physics, Beijing Normal University, Beijing, 100875, China\\
 $^2$ Institute of Theoretical Physics, Zhanjiang Normal University, Zhanjiang, 524048, China\\
  }

\begin{abstract}
 To provide an analytic verification of the nature of phase transition at the critical point of $P-V$ criticality, the original expressions of Ehrenfest equations have been introduced directly. By treating the cosmological constant and its conjugate quantity as thermodynamic pressure and volume respectively, we carry out analytical check of classical Ehrenfest equations. To show that our approach is universal, we investigate not only higher-dimensional charged AdS black holes, but also rotating AdS black holes. Not only are the examples of Einstein gravity shown, but also the example of modified gravity is presented for Gauss-Bonnet AdS black holes. The specific heat at constant pressure $C_P$, the volume expansion coefficient $\alpha$ and the isothermal compressibility coefficient $\kappa_T$ are found to diverge exactly at the critical point. It has been verified that both Ehrenfest equations hold at the critical point of $P-V$ criticality in the extended phase spaces of AdS black holes. So the nature of the critical point of $P-V$ criticality of AdS black holes has been demonstrated analytically to be a second-order phase transition. These results are consistent with the nature of liquid-gas phase transition at the critical point. In this sense, our research would deepen the understanding of the relations of AdS black holes and liquid-gas systems. Moreover, our successful approaches to introduce the original expressions of Erhenfest equations directly into black hole phase transition research demonstrate again that black hole thermodynamics is closely related to classical thermodynamics, which allows us to borrow techniques from classical thermodynamics to investigate the thermodynamics of black holes.
\end{abstract}

  \maketitle

\section{Introduction}
     Black hole thermodynamics has been an exciting and challenging topic in theoretical physics since the pioneer research in which black holes were identified as thermodynamic objects with both temperature and entropy~\cite{Bekenstein,Hawking1}. Phase transition, a fascinating phenomenon in classical thermodynamics, has aroused the attention of researchers. In 1983, Hawking and Page demonstrated that there exists a phase transition between the Schwarzschild AdS black hole and thermal AdS space \cite{Hawking2}, opening an era for phase transition research of AdS black holes. It is believed that phase transitions of AdS black holes are of great importance for they may serve as a novel way to investigate phase transitions in the dual field theories\cite{Witten}.

          In classical thermodynamics, one can classify phase transitions as first order or higher order
transitions by utilizing Clausius-Clapeyron-Ehrenfest equations. For a first order transition the Clausius-Clapeyron
equation is satisfied while for a second order transition Ehrenfest equations are satisfied. Recently, Banerjee et al. developed an innovative scheme based on the beautiful analogy of Ehrenfest¡¯s relations by considering the analogy ($V\leftrightarrow Q,P\leftrightarrow-\Phi$) between the thermodynamic state variables and black hole parameters~\cite{Banerjee1}-\cite{Banerjee6}. To study phase transitions in black holes, they considered the black holes as grand-canonical ensembles and performed a detailed analysis of the analogy of Ehrenfest¡¯s relations using both analytical and graphical techniques.

On the other hand, it is found that phase transitions of charged AdS black holes behaves much like the liquid-gas system. The observation of this phenomenon was first made by Chamblin et al.~\cite{Chamblin1,Chamblin2}. They found that the critical behavior of Reissner-Nordstrom-AdS (RN-AdS) black hole is analogous to the Van der Waals liquid-gas phase transition. Recently, there has been increasing attention of considering the variation of the cosmological constant in the first law of black hole thermodynamics~\cite{Caldarelli}-\cite{Lu}. Treating the cosmological constant as pressure and conjugate variable as volume, the
analogy between a charged AdS black hole and the liquid-gas system has been completed~\cite{Dolan2}. Moreover, it was argued that there are at least three reasons to consider variation of the cosmological constant~\cite{Kubiznak}. Firstly, there exist some more fundamental theories where physical
constants (e.g. Yukawa coupling, gauge coupling constants, Newton¡¯s constant, or cosmological constant) can vary because they arise as vacuum expectation values. In such theories, it is natural to consider the variations of physical constants in the first law. Secondly, only when one include the variations of cosmological constant in the first law, can the first law of black hole thermodynamics becomes consistent with the Smarr relations. Thirdly, the black hole mass is identified as enthalpy rather than internal energy if one considers the variation of cosmological constant in the first law. Based on these considerations, Kubiz\v{n}\'{a}k et al.~\cite{Kubiznak} investigated the $P-V$ criticality of charged AdS black holes by treating the cosmological constant as thermodynamic pressure and its conjugate quantity as thermodynamic volume. Their work has been elaborated subsequently~\cite{Gunasekaran}-\cite{Altamirano2} and $P-V$ criticality in the extended phase space (including pressure and volume as thermodynamic variables) of black holes has led a new trend of phase transition research.

 Recently, we have successfully introduced the original expressions of Erhenfest equations to prove analytically that RN-AdS black holes in the extened phase space undergo second order phase transition at the critical point~\cite{Wenbiao1}. To show that our approach is universal, we would like to elaborate on our recent research in this paper. We would not only investigate higher dimensional charged AdS black holes, but also investigate the rotating AdS black holes. We would not only show the examples of Einstein gravity, but also present the example of modified gravity by studying the Gauss-Bonnet AdS black holes. Refs.~\cite{Gunasekaran} and \cite{Cai98} calculated the critical exponents for the above black holes and showed that they are in the same university class as the van der Waals
 gas, which is known to have a second order phase transition at the critical point. So our paper would serve as both an independent and analytic verification of the nature of phase transition at the critical point of the above black holes.

The organization of this paper is as follows. In Sec.\ref{Sec2}, the original expressions of Ehrenfest equations will be introduced and an analytical check of both equations would be carried out at the critical point of $P-V$ criticality of higher dimensional charged AdS black holes. In Sec.\ref{Sec3}, an analytical check of Ehrenfest equations would be carried out at the critical point of $P-V$ criticality of rotating AdS black holes. In Sec.\ref{Sec4}, an analytical check of Ehrenfest equations would be carried out at the critical point of $P-V$ criticality of Gauss-Bonnet AdS black holes. In the end, a brief conclusion will be drawn in Sec.\ref {Sec5}. Note that notations of physical quantities have independent physical meanings in each section. For example, $T$ stands for the Hawking temperature of higher dimensional charged AdS black holes in Sec.\ref{Sec2} while it stands for the Hawking temperature of rotating AdS black holes in Sec.\ref{Sec3}.
\section{Analytical check of the classical Ehrenfest equations in the extended phase space of higher dimensional charged AdS black holes}

\label{Sec2} To start with, we would like to briefly review the
$P-V$ criticality of higher dimensional charged AdS black holes in the extended phase space.

The bulk action and the solution of a spherical charged AdS black hole in higher dimensions have been reviewed in Ref.~\cite{Gunasekaran} as
\begin{eqnarray}
I_{EM}&=&-\frac{1}{16\pi}\int_M d^dx\sqrt{-g}\left(R-F^2+\frac{(d-1)(d-2)}{l^2}\right),\label{1}
\\
ds^2&=&-fdt^2+\frac{dr^2}{f}+r^2d\Omega_{d-2}^2, \nonumber
\\
F&=&dA, \; \; A=-\sqrt{\frac{d-2}{2(d-3)}}\frac{q}{r^{d-3}}dt,\label{2}
\end{eqnarray}%
where
\begin{equation}
f=1-\frac{m}{r^{d-3}}+\frac{q^2}{r^{2(d-3)}}+\frac{r^2}{l^2}.\label{3}
\end{equation}%
The ADM mass and the electric charge has been identified as ~\cite{Chamblin1}
\begin{eqnarray}
M&=&\frac{\omega_{d-2}(d-2)}{16\pi}m,\label{4}
\\
Q&=&\frac{\omega_{d-2}\sqrt{2(d-2)(d-3)}}{8\pi}q,\label{5}
\end{eqnarray}%
where the volume of the unit $d$-sphere $\omega_d$ can be expressed as
\begin{equation}
\omega_d=\frac{2\pi^{\frac{d+1}{2}}}{\Gamma(\frac{d+1}{2})}.\label{6}
\end{equation}%
The corresponding Hawking temperature, entropy and electric potential have been reviewed in Ref.~\cite{Gunasekaran} as
\begin{eqnarray}
T&=&\frac{f'(r_+)}{4\pi}=\frac{d-3}{4\pi r_+}\left(1-\frac{q^2}{r_+^{2(d-3)}}+\frac{d-1}{d-3} \frac{r_+^2}{l^2}\right),\label{7}
\\
S&=&\frac{\omega_{d-2}r_+^{d-2}}{4}, \label{8}
\\
\Phi&=&\sqrt{\frac{d-2}{2(d-3)}}\frac{q}{r_+^{d-3}}. \label{9}
\end{eqnarray}%

The most creative point of $P-V$ criticality is by treating the cosmological constant as thermodynamic pressure and its conjugate quantity as
thermodynamic volume one can actually compare the same physical quantities rather than an analogy. As for charged AdS black holes in higher dimensions, the pressure~\cite{Gunasekaran} and thermodynamic volume~\cite{Cvetic} of RN-AdS black hole have been identified as
\begin{eqnarray}
P&=&-\frac{\Lambda}{8\pi}=\frac{(d-1)(d-2)}{16\pi l^2},\label{10}
\\
V&=&\frac{\omega_{d-2}r_+^{d-1}}{d-1}. \label{11}
\end{eqnarray}%
With the newly introduced thermodynamic pressure and volume, the first law of black hole thermodynamics~\cite{Dolan2} and the corresponding Smarr relation in the extended phase space take new forms as
\begin{eqnarray}
dM&=&TdS+\Phi dQ+VdP,\label{12}
\\
M&=&\frac{d-2}{d-3}TS+\Phi Q-\frac{2}{d-3}VP. \label{13}
\end{eqnarray}%
The critical point was also obtained in Ref.~\cite{Gunasekaran} as
\begin{eqnarray}
v_c&=&\frac{1}{\kappa}\left[q^2(d-2)(2d-5)\right]^{1/[2(d-3)]},
\nonumber
\\
T_c&=&\frac{(d-3)^2}{\pi \kappa v_c(2d-5)},
\nonumber
\\
P_c&=&\frac{(d-3)^2}{16\pi \kappa^2v_c^2},\label{14}
\end{eqnarray}%
where the specific volume $v$ is related to the horizon radius $r_+$ by
\begin{equation}
r_+=\kappa v, \;\kappa=\frac{d-2}{4}.\label{15}
\end{equation}

In classical thermodynamics, one can classify phase transitions as first order or higher order
transitions by utilizing Clausius-Clapeyron-Ehrenfest equations. For a first order phase transition the Clausius-Clapeyron
equation is satisfied, while for a second order phase transition, Ehrenfest equations are satisfied.

  The classical Ehrenfest equations can be found in any textbook of classical thermodynamics as
\begin{eqnarray}
(\frac{\partial P}{\partial
T})_S&=&\frac{C_{P_2}-C_{P_1}}{VT(\alpha_2-\alpha_1)}=\frac{\Delta
C_P}{VT\Delta \alpha},\label{18}\\
(\frac{\partial P}{\partial
T})_V&=&\frac{\alpha_2-\alpha_1}{\kappa_{T_2}-\kappa_{T_1}}=\frac{\Delta
\alpha}{\Delta\kappa_T},\label{19}
\end{eqnarray}%
 where $\alpha=\frac{1}{V}(\frac{\partial V}{\partial T})_P$ is
volume expansion coefficient and $\kappa_T=-\frac{1}{V}(\frac{\partial V}{\partial P})_T$ is isothermal compressibility coefficient.

 Now let's embark on checking the validity of Ehrenfest¡¯s equations
(\ref{18})-(\ref{19}) at the critical point of $P-V$ criticality in the extended phase space of higher dimensional charged AdS black hole. To start with, we would like to calculate the relevant quantities in (\ref{18})-(\ref{19}).

 Substituting Eqs.(\ref{8}) and (\ref{10}) into Eq.(\ref{7}), we obtain
  \begin{equation}
T=\frac{(d-3)}{4\pi}\left[\left(\frac{4S}{\omega_{d-2}}\right)^{-\frac{1}{d-2}}-q^2\left(\frac{4S}{\omega_{d-2}}\right)^{\frac{5-2d}{d-2}}+\frac{16\pi P\left(\frac{4S}{\omega_{d-2}}\right)^{\frac{1}{d-2}}}{(d-3)(d-2)}\right].\label{20}
\end{equation}%
 When $d=4$, Eq.(\ref{20}) can be simplified into
 \begin{equation}
T=\frac{1}{4\sqrt{\pi S}}(1+8PS-\frac{\pi Q^2}{S}),\label{21}
\end{equation}%
which is consistent with the results of four dimensional RN-AdS black hole in Ref.~\cite{Kubiznak}.

Utilizing Eqs.(\ref{8}), (\ref{11}) and (\ref{20}), the specific heat at constant pressure, volume expansion coefficient and isothermal compressibility coefficient can be derived as
\begin{eqnarray}
C_P&=&T(\frac{\partial S}{\partial T})_P=\frac{(d-2)S\bigg[1-q^2\left(\frac{4S}{\omega_{d-2}}\right)^{\frac{6-2d}{d-2}}+\frac{16P\pi\left(\frac{4S}{\omega_{d-2}}\right)^{\frac{2}{d-2}}}{(d-3)(d-2)}\bigg]}{-1+(2d-5)q^2\left(\frac{4S}{\omega_{d-2}}\right)^{\frac{6-2d}{d-2}}+\frac{16P\pi\left(\frac{4S}{\omega_{d-2}}\right)^{\frac{2}{d-2}}}{(d-3)(d-2)}},\label{22}\\
\alpha&=&\frac{1}{V}(\frac{\partial V}{\partial
T})_P=\frac{4\pi(\frac{d-1}{d-3})\left(\frac{4S}{\omega_{d-2}}\right)^{\frac{1}{d-2}}}{-1+(2d-5)q^2\left(\frac{4S}{\omega_{d-2}}\right)^{\frac{6-2d}{d-2}}+\frac{16P\pi\left(\frac{4S}{\omega_{d-2}}\right)^{\frac{2}{d-2}}}{(d-3)(d-2)}},\label{23}\\
\kappa_T&=&-\frac{1}{V}(\frac{\partial V}{\partial P})_T=\frac{16\pi\frac{(d-1)}{(d-2)(d-3)}\left(\frac{4S}{\omega_{d-2}}\right)^{\frac{2}{d-2}}}{-1+(2d-5)q^2\left(\frac{4S}{\omega_{d-2}}\right)^{\frac{6-2d}{d-2}}+\frac{16P\pi\left(\frac{4S}{\omega_{d-2}}\right)^{\frac{2}{d-2}}}{(d-3)(d-2)}}.\label{24}
\end{eqnarray}%
In the derivation of Eq.(\ref{24}), we have utilized the thermodynamic identity as
\begin{equation}
(\frac{\partial V}{\partial P})_T(\frac{\partial P}{\partial
T})_V(\frac{\partial T}{\partial V})_P=-1.\label{25}
\end{equation}%
It is quite interesting to note that $C_P,\alpha,\kappa_T$ share
exactly the same factor in their denominators.
Utilizing Eqs.(\ref{8}), (\ref{14}) and (\ref{15}), we obtain that
\begin{eqnarray}
S_c&=&\pi\left[(d-2)(2d-5)q^2\right]^{\frac{d-2}{2d-6}},\label{26}
\\
P_c&=&\frac{(d-3)^2}{16\pi}\left[(d-2)(2d-5)q^2\right]^{\frac{-2}{2d-6}}.\label{27}
\end{eqnarray}%
With Eqs.(\ref{26}) and (\ref{27}),we can derive that
\begin{equation}
-1+(2d-5)q^2\left(\frac{4S_c}{\omega_{d-2}}\right)^{\frac{6-2d}{d-2}}+\frac{16P_c\pi\left(\frac{4S_c}{\omega_{d-2}}\right)^{\frac{2}{d-2}}}{(d-3)(d-2)}=0.\label{28}
\end{equation}%
Eq.(\ref{28}) implies that $C_P,\alpha,\kappa_T$ diverge at the critical
point.

From the definition of volume expansion coefficient $\alpha$, we obtain
\begin{equation}
V\alpha=(\frac{\partial V}{\partial T})_P=(\frac{\partial
V}{\partial S})_P(\frac{\partial S}{\partial
T})_P=(\frac{\partial V}{\partial S})_P(\frac{C_P
}{T}).\label{29}
\end{equation}%
Then the R.H.S of Eq.(\ref{18}) can be transformed into
\begin{equation}
\frac{\Delta C_P}{TV\Delta \alpha}=[(\frac{\partial S}{\partial
V})_P]_c.\label{30}
\end{equation}%
Note that the footnote "c" denotes the values of physical quantities at the critical point in this paper.
Utilizing Eqs.(\ref{8}), (\ref{11}) and (\ref{30}), we obtain
\begin{equation}
\frac{\Delta C_P}{TV\Delta \alpha}=\frac{1}{4}(d-2)\left(\frac{4S_c}{\omega_{d-2}}\right)^{-\frac{1}{d-2}}.\label{31}
\end{equation}%
Utilizing Eq.(\ref{20}), the L.H.S of Eq.(\ref{18}) can be derived as
\begin{equation}
[(\frac{\partial P}{\partial T})_S]_c=\frac{1}{4}(d-2)\left(\frac{4S_c}{\omega_{d-2}}\right)^{-\frac{1}{d-2}}.\label{32}
\end{equation}%
    From Eqs.(\ref{31}) and (\ref{32}), we can draw the conclusion that the first equation of Erhenfest equations is valid at the
    critical point.

    Utilizing Eqs.(\ref{8}), (\ref{11}) and (\ref{20}), the L.H.S of Eq.(\ref{19}) can be obtained as
\begin{equation}
[(\frac{\partial P}{\partial T})_V]_c=\frac{1}{4}(d-2)\left(\frac{4S_c}{\omega_{d-2}}\right)^{-\frac{1}{d-2}}.\label{33}
\end{equation}%

Utilizing the definitions of isothermal compressibility coefficient $\kappa_T$ and volume expansion coefficient $\alpha$, we can derive that
\begin{equation}
V\kappa_T=-(\frac{\partial V}{\partial P})_T=(\frac{\partial
T}{\partial P})_V(\frac{\partial V}{\partial
T})_P=(\frac{\partial T}{\partial P})_VV\alpha,\label{34}
\end{equation}%
from which we can calculate the R.H.S of Eq.(\ref{19}) and get
\begin{equation}
\frac{\Delta \alpha}{\Delta \kappa_T}=[(\frac{\partial
P}{\partial T})_V]_c=\frac{1}{4}(d-2)\left(\frac{4S_c}{\omega_{d-2}}\right)^{-\frac{1}{d-2}}.\label{35}
\end{equation}%
In the derivation of Eq.(\ref{34}), we have utilized the thermodynamic identity-Eq.(\ref{25}) again. Eq.(\ref{35}) reveal the validity of the second equation of Ehrenfest equations.

                So far, we have proved that both the Ehrenfest equations are correct at the critical point. Utilizing Eqs.(\ref{31}) and (\ref{35}), the Prigogine-Defay(PD)
                ratio can be calculated as
\begin{equation}
\Pi=\frac{\Delta C_P \Delta \kappa_T}{TV(\Delta
\alpha)^2}=1.\label{36}
\end{equation}%
The definition of PD ratio was presented by Prigogine and Defay~\cite{Prigogine} and reviewed in Ref.~\cite{Moynihan}. The second Ehrenfest equation is not always satisfied and the PD ratio can be used to measure the deviation from the second Ehrenfest equation~\cite{Banerjee1,Banerjee4}. It was discovered that for a second order phase transition it is equal to one~\cite{Banerjee1} while for a glassy phase transition its value varies from 2 to 5~\cite{Banerjee1,Jackle,Nieuwenhuizen1}. It was argued that~\cite{Banerjee1} for a normal second order phase transition, first order derivatives of Gibb¡¯s free energy are continuous at the phase transition point and hence one can equate the left hand sides of two Ehrenfest equations to get the result $\Pi=1$.

                Eq.(\ref{36}) and the validity of Ehrenfest equations prove that the phase transition at the critical point of $P-V$ criticality in the extended phase space of higher dimensional charged AdS black hole is a second order transition. This result is consistent with the nature of liquid-gas phase transition at the critical point, hence deepening the understanding of the analogy of charged AdS black holes and liquid-gas system. It is necessary to emphasize that although $C_P,\alpha,\kappa_T$ diverge at the critical point, they can cancel each other and allow the R.H.S of  Eqs.(\ref{18}) and  (\ref{19})  to be finite.

\section{Analytical check of the classical Ehrenfest equations in the extended phase space of rotating AdS black holes}
\label{Sec3} In this section, we would like to carry out an analytical check of Ehrenfest equations at the critical point of $P-V$ criticality of rotating AdS black holes.

To begin with, we would briefly review the
$P-V$ criticality of rotating AdS black holes in the extended phase space. The famous Kerr-Newman-AdS metric, which describes the charged AdS rotating black hole solution, has been proposed as
\begin{equation}
ds^2=-\frac{\Delta}{\rho^2}\left[dt-\frac{a\sin\theta^2}{\Xi}d\varphi\right]^2+\frac{\rho^2}{\Delta}dr^2+\frac{\rho^2}{\Delta_\theta}d\theta^2+\frac{\Delta_\theta \sin^2\theta}{\rho^2}\left[adt-\frac{r^2+a^2}{\Xi}d\varphi\right]^2,\label{301}
\end{equation}%
where
\begin{eqnarray}
\rho^2&=&r^2+a^2\cos^2\theta,\;\Xi=1-\frac{a^2}{l^2},\;\Delta_\theta=1-\frac{a^2}{l^2}\cos^2\theta,
\nonumber
\\
\Delta&=&(r^2+a^2)(1+\frac{r^2}{l^2})-2mr+q^2
.\label{302}
\end{eqnarray}%
The mass $M$, the electric charge $Q$ and the angular momentum $J$ are related to the parameters as
\begin{equation}
M=\frac{m}{\Xi^2},\;Q=\frac{q}{\Xi},\;J=\frac{am}{\Xi^2}.\label{303}
\end{equation}%
The corresponding Hawking temperature, entropy, electric potential and angular velocity have been reviewed in Ref.~\cite{Gunasekaran} as
\begin{eqnarray}
T&=&\frac{r_+(1+\frac{a^2}{l^2}+\frac{3r_+^2}{l^2}-\frac{a^2+q^2}{r_+^2})}{4\pi (r_+^2+a^2)},\label{304}
\\
S&=&\frac{\pi(r_+^2+a^2)}{\Xi}, \label{305}
\\
\Phi&=&\frac{qr_+}{r_+^2+a^2}, \label{306}
\\
\Omega&=&\frac{a(l^2+r_+^2)}{l^2(r_+^2+a^2)}. \label{307}
\end{eqnarray}%

Treating the cosmological constant as thermodynamic pressure and its conjugate quantity as thermodynamic volume, the pressure~\cite{Gunasekaran} and thermodynamic volume~\cite{Dolan3,Cvetic} of Kerr-Newman-AdS black hole have been identified as
\begin{eqnarray}
P&=&\frac{3}{8\pi l^2},\label{308}
\\
V&=&\frac{2\pi\left[(r_+^2+a^2)(2r_+^2l^2+a^2l^2-r_+^2a^2)+l^2q^2a^2\right]}{3l^2\Xi^2r_+}. \label{309}
\end{eqnarray}%
And the first law of black hole thermodynamics~\cite{Dolan2} take new forms as
\begin{equation}
dM=TdS+VdP+\Omega dJ+\Phi dQ.\label{310}
\end{equation}%
When the charge $Q=0$, the critical point was also obtained in Ref.~\cite{Gunasekaran} as
\begin{equation}
v_c=2\times90^{1/4}\sqrt{J},\;T_c=\frac{90^{3/4}}{225\pi \sqrt{J}},\;P_c=\frac{1}{12\sqrt{90}\pi J}.\label{311}
\end{equation}%

 To start with, we would like to calculate the relevant quantities in (\ref{18})-(\ref{19}) for $P-V$ criticality in the extended phase space of rotating AdS black holes. It is more convenient to reexpress the mass (the enthalpy) and the temperature in terms of $S$, $J$, $P$ and $Q$. Such expressions can be found in Ref.~\cite{Dolan2} as
\begin{eqnarray}
H&=&\frac{1}{2}\sqrt{\frac{\left(S+\pi Q^2+\frac{8PS^2}{3}\right)^2+4\pi^2\left(1+\frac{8PS}{3}\right)J^2}{\pi S}},\label{312}
\\
T&=&\frac{1}{8\pi H}\left[\left(1+\frac{\pi Q^2}{S}+\frac{8PS}{3}\right)\left(1-\frac{\pi Q^2}{S}+8PS\right)-4\pi^2(\frac{J}{S})^2\right],\label{313}
\\
V&=&\frac{2}{3\pi H}\left[S\left(S+\pi Q^2+\frac{8PS^2}{3}\right)+2\pi^2J^2\right].\label{314}
\end{eqnarray}%
Note that the mass has been interpreted as the enthalpy when considering the variation of cosmological constant in the first law.

Utilizing Eqs.(\ref{313}) and (\ref{314}), the specific heat at constant pressure, volume expansion coefficient and isothermal compressibility coefficient can be derived as
\begin{eqnarray}
C_P=T(\frac{\partial S}{\partial T})_P&=&\frac{2S}{B(S,P,J)}\times\left[12J^2\pi^2(3+8PS)+(3\pi Q^2+3S+8PS^2)^2\right]
\nonumber
\\
&\;& \times \left[-12J^2\pi^2-(\pi Q^2-S-8PS^2)(3\pi Q^2+3S+8PS^2)\right],\label{315}\\
\alpha=\frac{1}{V}(\frac{\partial V}{\partial
T})_P&=&\sqrt{12J^2\pi^2(3+8PS)+(3\pi Q^2+3S+8PS^2)^2}\times\{S(3S+3\pi Q^2+8PS^2)^3
\nonumber
\\
&\;&+72J^4\pi^4+6J^2\pi^2\left[3\pi ^2Q^4+3S^2(3+8PS)^2+2\pi Q^2S(9+8PS)\right]\}\nonumber
\\
&\;& \times \frac{12\sqrt{\pi}S^{3/2}}{B(S,P,J)\times\left[6J^2\pi^2+S(3\pi Q^2+3S+8PS^2)\right]}
,\label{316}\\
\kappa_T=-\frac{1}{V}(\frac{\partial V}{\partial P})_T&=&\frac{24S}{B(S,P,J)\times\left[6J^2\pi^2+S(3\pi Q^2+3S+8PS^2)\right]}
\nonumber
\\
&\;& \times \{576J^6\pi^6+S^2(3\pi Q^2+3S+8PS^2)^4
\nonumber
\\
&\;& +24J^2\pi^2S(3\pi Q^2+3S+8PS^2)\left[3\pi^2 Q^4+4\pi Q^2S+S^2(3+8PS)^2\right]
\nonumber
\\
&\;&+48J^4\pi^4\left[3\pi^2 Q^4+18\pi Q^2S+S^2(3+8PS)(5+8PS)\right]\}
,\label{317}
\end{eqnarray}%
where
\begin{eqnarray}
B(S,P,J)&=&144J^4\pi^4(9+32PS)+(3\pi Q^2-S+8PS^2)(3\pi Q^2+3S+8PS^2)^3
\nonumber
\\
&\;&+24J^2\pi^2\left[36\pi Q^2S(1+4PS)+S^2(3+8PS)^2(3+16PS)+3\pi^2Q^4(9+16PS)\right]
.\label{318}
\end{eqnarray}%
In the derivation of Eq.(\ref{317}), we have utilized the formula as follow
\begin{equation}
(\frac{\partial V}{\partial P})_T=(\frac{\partial V}{\partial P})_S+(\frac{\partial V}{\partial S})_P(\frac{\partial S}{\partial P})_T.\label{319}
\end{equation}%
It is quite interesting to note that $C_P,\alpha,\kappa_T$ share
the same factor $B(S,P,J)$ in their denominators. When the charge $Q=0$, by utilizing Eq.(\ref{311}), we can obtain that
\begin{eqnarray}
B(S_c,P_c,J_c)&=&144J_c^4\pi^4(9+32P_cS_c)+(-S_c+8P_cS_c^2)(3S_c+8P_cS_c^2)^3+24\pi^2 J_c^2S_c^2(3+8P_cS_c)^2(3+16P_cS_c)
\nonumber
\\
&=&0
,\label{320}
\end{eqnarray}%
which implies that $C_P,\alpha,\kappa_T$ may diverge at the critical
point.

Now let's embark on checking the validity of Ehrenfest¡¯s equations
(\ref{18})-(\ref{19}) at the critical point. From the definition of volume expansion coefficient $\alpha$, we obtain
\begin{equation}
V\alpha=(\frac{\partial V}{\partial T})_P=(\frac{\partial
V}{\partial S})_P(\frac{\partial S}{\partial
T})_P=(\frac{\partial V}{\partial S})_P(\frac{C_P
}{T}).\label{321}
\end{equation}%
Then the R.H.S of Eq.(\ref{18}) can be transformed into
\begin{equation}
\frac{\Delta C_P}{TV\Delta \alpha}=[(\frac{\partial S}{\partial
V})_P]_c.\label{322}
\end{equation}%
Utilizing Eqs.(\ref{305}), (\ref{309}) and (\ref{322}), we obtain
\begin{equation}
\frac{\Delta C_P}{TV\Delta \alpha}=\frac{\sqrt{\pi S_c}\left[12J_c^2\pi^2(3+8P_cS_c)+(3\pi Q^2+3S_c+8P_cS_c^2)^2\right]^{3/2}
}{C(S_c,P_c,J_c)},\label{323}
\end{equation}%
where
\begin{equation}
C(S_c,P_c,J_c)=144J_c^4\pi^4+2S_c(3\pi Q^2+3S_c+8P_cS_c^2)^3+12J_c^2\pi ^2\left[3\pi^2Q^4+3S_c^2(3+8P_cS_c)^2+2\pi Q^2S_c(9+8P_cS_c)\right].\label{324}
\end{equation}
Utilizing Eq.(\ref{313}), the L.H.S of Eq.(\ref{18}) can be derived as
\begin{equation}
[(\frac{\partial P}{\partial T})_S]_c=\frac{\sqrt{\pi S_c}\left[12J_c^2\pi^2(3+8P_cS_c)+(3\pi Q^2+3S_c+8P_cS_c^2)^2\right]^{3/2}
}{C(S_c,P_c,J_c)}.\label{325}
\end{equation}
    From Eqs.(\ref{323}) and (\ref{325}), we can draw the conclusion that the first equation of Erhenfest equations is valid at the
    critical point.

    Utilizing Eqs.(\ref{316}) and (\ref{317}), the L.H.S of Eq.(\ref{19}) can be obtained as
\begin{eqnarray}
[(\frac{\partial P}{\partial T})_V]_c&=&\frac{\sqrt{\pi S_c\left[12J_c^2\pi^2(3+8P_cS_c)+(3\pi Q^2+3S_c+8P_cS_c^2)^2\right]}
}{D(S_c,P_c,J_c)}
\nonumber
\\
&\;& \times  \{72J_c^4\pi ^4+S_c\left(3\pi Q^2+3S_c+8P_cS_c^2\right)^3
\nonumber
\\
&\;&+6J_c^2\pi ^2\left[3\pi ^2Q^4+3S_c^2(3+8P_cS_c)^2+2\pi Q^2S_c(9+8P_cS_c)\right]\}
,\label{326}
\end{eqnarray}%
where
\begin{eqnarray}
D(S_c,P_c,J_c)&=&1152J_c^6\pi^6+S_c^2(3\pi Q^2+3S_c+8P_cS_c^2)^4
\nonumber
\\
&\;&+48J_c^2\pi ^2S_c(3\pi Q^2+3S_c+8P_cS_c^2)\left[3\pi^2Q^4+4\pi Q^2S_c+S_c^2(3+8P_cS_c)^2\right]
\nonumber
\\
&\;&+96J_c^4\pi ^4\left[3\pi^2Q^4+18\pi Q^2S_c+S_c^2(3+8P_cS_c)(5+8P_cS_c)\right].\label{327}
\end{eqnarray}
In the derivation of Eqs.(\ref{326}), we have utilized the thermodynamic identity Eq.(\ref{25}) again.

Utilizing the definitions of isothermal compressibility coefficient $\kappa_T$ and volume expansion coefficient $\alpha$, we can derive that
\begin{equation}
V\kappa_T=-(\frac{\partial V}{\partial P})_T=(\frac{\partial
T}{\partial P})_V(\frac{\partial V}{\partial
T})_P=(\frac{\partial T}{\partial P})_VV\alpha,\label{328}
\end{equation}%
from which we can calculate the R.H.S of Eq.(\ref{19}) and get
\begin{eqnarray}
\frac{\Delta \alpha}{\Delta \kappa_T}=[(\frac{\partial
P}{\partial T})_V]_c&=&\frac{\sqrt{\pi S_c\left[12J_c^2\pi^2(3+8P_cS_c)+(3\pi Q^2+3S_c+8P_cS_c^2)^2\right]}
}{D(S_c,P_c,J_c)}
\nonumber
\\
&\;& \times \{72J_c^4\pi ^4+S_c\left(3\pi Q^2+3S_c+8P_cS_c^2\right)^3
\nonumber
\\
&\;&+6J_c^2\pi ^2\left[3\pi ^2Q^4+3S_c^2(3+8P_cS_c)^2+2\pi Q^2S_c(9+8P_cS_c)\right]\}.\label{329}
\end{eqnarray}%
In the derivation of Eqs.(\ref{328}), we have utilized the thermodynamic identity Eq.(\ref{25}) again.

 Eq.(\ref{329}) reveal the validity of the second equation of Ehrenfest equations. So far, we have proved that both the Ehrenfest equations are correct at the critical point.

 Utilizing Eqs.(\ref{323}) and (\ref{329}), one can obtain
 \begin{eqnarray}
\frac{\Delta C_P}{TV\Delta \alpha}-\frac{\Delta \alpha}{\Delta \kappa_T}&=&\frac{24J_c^2\pi^3(J_c^2\pi +Q^2S_c)\times B(S_c,P_c,J_c)
}{C(S_c,P_c,J_c)\times D(S_c,P_c,J_c)}
\nonumber
\\
&\;& \times \sqrt{\pi S_c\left[12J_c^2\pi^2(3+8P_cS_c)+(3\pi Q^2+3S_c+8P_cS_c^2)^2\right]}
.\label{330}
\end{eqnarray}%
Keeping the fact that $B(S_c,P_c,J_c)$ is equal to zero at the critical point in mind, one can easily get
 \begin{equation}
\frac{\Delta C_P}{TV\Delta \alpha}-\frac{\Delta \alpha}{\Delta \kappa_T}=0.\label{331}
\end{equation}%

                So the $PD$ ratio can be derived as
\begin{equation}
\Pi=\frac{\Delta C_P \Delta \kappa_T}{TV(\Delta
\alpha)^2}=1.\label{332}
\end{equation}%

                Eq.(\ref{332}) and the validity of Ehrenfest equations prove that the phase transition at the critical point of P-V criticality in the extended phase space of rotating AdS black holes is a second order transition.

\section{Analytical check of the classical Ehrenfest equations in the extended phase space of Gauss-Bonnet AdS black holes}
\label{Sec4}
In this section, we would like to carry out an analytical check of Ehrenfest equations at the critical point of $P-V$ criticality of Gauss-Bonnet AdS black holes.

To start with, we would briefly review the $P-V$ criticality of Gauss-Bonnet AdS black holes in the extended phase space. The action and the solution of $d$-dimensional Einstein-Maxwell theory with a Gauss-Bonnet term and a cosmological constant have been reviewed in Ref.~\cite{Cai98} as
\begin{eqnarray}
S&=&\frac{1}{16\pi}\int d^dx\sqrt{-g}\left[R-2\Lambda+\alpha_{GB}(R_{\mu\nu\gamma\delta}R^{\mu\nu\gamma\delta}-4R_{\mu\nu}R^{\mu\nu}+R^2)-4\pi F_{\mu\nu}F
^{\mu\nu}\right],\label{401}
\\
ds^2&=&-f(r)dt^2+f(r)^{-1}dr^2+r^2h_{ij}dx^idx^j
,\label{402}
\end{eqnarray}%
where
\begin{equation}
f(r)=k+\frac{r^2}{2\tilde{\alpha}}\left(1-\sqrt{1+\frac{64\pi \tilde{\alpha}M}{(d-2)\sum_kr^{d-1}}-\frac{2\tilde{\alpha}Q^2}{(d-2)(d-3)r^{2d-4}}-\frac{64\pi \tilde{\alpha}P}{(d-1)(d-2)}}\,\right).\label{403}
\end{equation}%
Solving the equation $f(r)=0$ for the largest root, one can obtain the horizon radius $r_h$. And the mass can be expressed as
\begin{equation}
M=\frac{(d-2)\sum_kr_h^{d-3}}{16\pi}\left(k+\frac{k^2\tilde{\alpha}}{r_h^2}+\frac{16\pi P r_h^2}{(d-1)(d-2)}\right)+\frac{\sum_kQ^2}{8\pi (d-3)r_h^{d-3}}.\label{404}
\end{equation}%

The corresponding Hawking temperature, entropy and electric potential have been derived as~\cite{Cai98}
\begin{eqnarray}
T&=&\frac{f'(r_h)}{4\pi}=\frac{16\pi Pr_h^4/(d-2)+(d-3)kr_h^2+(d-5)k^2\tilde{\alpha}-\frac{2Q^2}{(d-2)r_h^{2d-8}}}{4\pi r_h(r_h^2+2k\tilde{\alpha})},\label{405}
\\
S&=&\frac{\sum_kr_h^{d-2}}{4}\left(1+\frac{2(d-2)\tilde{\alpha}k}{(d-4)r_h^2}\right), \label{406}
\\
\Phi&=&\frac{\sum_kQ}{4\pi (d-3)r_h^{d-3}}. \label{407}
\end{eqnarray}%

Treating the cosmological constant as thermodynamic pressure and its conjugate quantity as
thermodynamic volume, the pressure and thermodynamic volume of Gauss-Bonnet AdS black hole have been identified as~\cite{Cai98}
\begin{eqnarray}
P&=&-\frac{\Lambda}{8\pi}=\frac{(d-1)(d-2)}{16\pi l^2},\label{408}
\\
V&=&\frac{\sum_kr_h^{d-1}}{d-1}. \label{409}
\end{eqnarray}%
With the newly introduced thermodynamic pressure and volume, the first law of black hole thermodynamics and the corresponding Smarr relation in the extended phase space take new forms as~\cite{Cai98}
\begin{eqnarray}
dH&=&TdS+\Phi dQ+VdP+\mathcal{A} d\tilde{\alpha},\label{410}
\\
(d-3)H&=&(d-2)TS-2PV+2\mathcal{A}\tilde{\alpha}+(d-3)Q\Phi, \label{411}
\end{eqnarray}%
where
\begin{equation}
\mathcal{A}=\left(\frac{\partial H}{\partial \tilde{\alpha}}\right)_{S,Q,P}=\frac{(d-2)k^2\sum_k}{16\pi}r_h^{d-5}-\frac{(d-2)k\sum_kT}{2(d-4)}r_h^{d-4} \label{412}
\end{equation}%
is the quantity conjugate to the Gauss-Bonnet coefficient $\tilde{\alpha}$.
When $Q=0,k=1,d=5$, the critical point was also obtained in Ref.~\cite{Cai98} as
\begin{equation}
r_{hc}=\sqrt{6\tilde{\alpha}},\;T_c=\frac{1}{\pi \sqrt{24\tilde{\alpha}}},\;P_c=\frac{1}{48\pi \tilde{\alpha}}.\label{413}
\end{equation}%

 To begin with, we would like to calculate the relevant quantities in (\ref{18})-(\ref{19}) for $P-V$ criticality in the extended phase space of Gauss-Bonnet AdS black hole. The corresponding Hawking temperature, entropy and electric potential have been derived as
\begin{eqnarray}
T&=&\frac{8\pi Pr_h^4+3r_h^2}{6\pi r_h(r_h^2+2\tilde{\alpha})},\label{414}
\\
S&=&\frac{\sum_1r_h^{3}}{4}\left(1+\frac{6\tilde{\alpha}}{r_h^2}\right), \label{415}
\\
V&=&\frac{\sum_1r_h^{4}}{4}. \label{416}
\end{eqnarray}%

Utilizing Eqs.(\ref{414}), (\ref{415}) and (\ref{416}), the specific heat at constant pressure, volume expansion coefficient and isothermal compressibility coefficient can be derived as
\begin{eqnarray}
C_P&=&T(\frac{\partial S}{\partial T})_P=\frac{3\sum_1r_h(3+8P\pi r_h^2)(r_h^2+2\tilde{\alpha})^2}{4\left[6\tilde{\alpha}-3r_h^2+8P\pi r_h^2(r_h^2+6\tilde{\alpha})\right]},\label{417}\\
\alpha&=&\frac{1}{V}(\frac{\partial V}{\partial
T})_P=\frac{24\pi (r_h^2+2\tilde{\alpha})^2}{r_h \left[6\tilde{\alpha}-3r_h^2+8P\pi r_h^2(r_h^2+6\tilde{\alpha})\right]},\label{418}\\
\kappa_T&=&-\frac{1}{V}(\frac{\partial V}{\partial P})_T=\frac{32\pi r_h^2(r_h^2+2\tilde{\alpha})}{6\tilde{\alpha}-3r_h^2+8P\pi r_h^2(r_h^2+6\tilde{\alpha})}.\label{419}
\end{eqnarray}%
In the derivation of Eqs.(\ref{419}), we have utilized the thermodynamic identity Eq.(\ref{25}) again.
It is quite interesting to note that $C_P,\alpha,\kappa_T$ share
the same factor in their denominators. Utilizing Eq.(\ref{413}), we obtain
\begin{equation}
6\tilde{\alpha}-3r_c^2+8P_c\pi r_c^2(r_c^2+6\tilde{\alpha})=0,\label{420}
\end{equation}%
implying that $C_P,\alpha,\kappa_T$ may diverge at the critical
point.

Now let's embark on checking the validity of Ehrenfest¡¯s equations
(\ref{18})-(\ref{19}) at the critical point. From the definition of volume expansion coefficient $\alpha$, we obtain
\begin{equation}
V\alpha=(\frac{\partial V}{\partial T})_P=(\frac{\partial
V}{\partial S})_P(\frac{\partial S}{\partial
T})_P=(\frac{\partial V}{\partial S})_P(\frac{C_P
}{T}).\label{421}
\end{equation}%
Then the R.H.S of Eq.(\ref{18}) can be transformed into
\begin{equation}
\frac{\Delta C_P}{TV\Delta \alpha}=[(\frac{\partial S}{\partial
V})_P]_c.\label{422}
\end{equation}%
Utilizing Eqs.(\ref{415}), (\ref{416}) and (\ref{422}), we obtain
\begin{equation}
\frac{\Delta C_P}{TV\Delta \alpha}=\frac{3(r_c^2+2\tilde{\alpha})
}{4r_c^3}.\label{423}
\end{equation}%
Utilizing Eqs.(\ref{408}), (\ref{414}) and (\ref{415}), the L.H.S of Eq.(\ref{18}) can be derived as
\begin{equation}
[(\frac{\partial P}{\partial T})_S]_c=\frac{3(r_c^2+2\tilde{\alpha})
}{4r_c^3}.\label{424}
\end{equation}%
    From Eqs.(\ref{423}) and (\ref{424}), we can draw the conclusion that the first equation of Erhenfest equations is valid at the
    critical point.

    Utilizing Eqs.(\ref{408}), (\ref{414}) and (\ref{416}), the L.H.S of Eq.(\ref{19}) can be obtained as
\begin{equation}
[(\frac{\partial P}{\partial T})_V]_c=\frac{3(r_c^2+2\tilde{\alpha})
}{4r_c^3}.\label{425}
\end{equation}%

Utilizing the definitions of isothermal compressibility coefficient $\kappa_T$ and volume expansion coefficient $\alpha$, we can derive that
\begin{equation}
V\kappa_T=-(\frac{\partial V}{\partial P})_T=(\frac{\partial
T}{\partial P})_V(\frac{\partial V}{\partial
T})_P=(\frac{\partial T}{\partial P})_VV\alpha,\label{426}
\end{equation}%
from which we can calculate the R.H.S of Eq.(\ref{19}) and get
\begin{equation}
\frac{\Delta \alpha}{\Delta \kappa_T}=[(\frac{\partial
P}{\partial T})_V]_c=\frac{3(r_c^2+2\tilde{\alpha})
}{4r_c^3}.\label{427}
\end{equation}%
In the derivation of Eq.(\ref{426}), we have utilized the thermodynamic identity-Eq.(\ref{25}) again. Eq.(\ref{427}) reveal the validity of the second equation of Ehrenfest equations.

                So far, we have proved that both the Ehrenfest equations are correct at the critical point. Utilizing Eqs.(\ref{423}) and (\ref{427}), the $PD$    ratio can be calculated as
\begin{equation}
\Pi=\frac{\Delta C_P \Delta \kappa_T}{TV(\Delta
\alpha)^2}=1.\label{428}
\end{equation}%

                Eq.(\ref{428}) and the validity of Ehrenfest equations prove that the phase transition at the critical point of $P-V$ criticality in the extended phase space of Gauss-Bonnet AdS black hole is a second order transition.

\section{Conclusions}
\label{Sec5}
In this paper, the original expressions of Ehrenfest equations have been introduced directly to provide an analytic verification of the nature of phase transition at the critical point of $P-V$ criticality. By treating the cosmological constant and its conjugate quantity as thermodynamic pressure and volume respectively, we carry out analytical check of classical Ehrenfest equations. We investigate three cases, higher dimensional charged AdS black holes, rotating AdS black holes and Gauss-Bonnet AdS black holes. For these three cases, the specific heat at constant pressure $C_P$, the volume expansion coefficient $\alpha$ and the isothermal compressibility coefficient $\kappa_T$ are calculated. It is shown that they share the same factor in their denominators and diverge exactly at the critical point. After analytical checks of Ehrenfest equations, both Ehrenfest equations have been verified to hold at the critical point of $P-V$ criticality in the extended phase spaces of higher dimensional charged AdS black holes, rotating AdS black holes and Gauss-Bonnet AdS black holes. Prigogine-Defay ratios are also calculated and it is shown that they are all exactly equal to one. So the nature of the critical point of $P-V$ criticality in the extended phase space of higher dimensional charged AdS black holes, rotating AdS black holes and Gauss-Bonnet AdS black holes have been demonstrated to be second order phase transition. Our approaches are not only valid for charged AdS black holes but also valid for rotating AdS black holes. Our approaches not only work for four-dimensional black holes but also works for higher-dimensional AdS black holes. Our approaches not only hold for black holes in Einstein gravity, but also hold for black holes in modified gravity. In this sense, our approaches are universal.

    Our results are consistent with the nature of liquid-gas phase transition at the critical point, demonstrating again that these black holes are in the same university class as the van der Waals
 gas. In this sense, our research would deepen the understanding of the relations of AdS black holes and liquid-gas system. Moreover, our successful approaches to introduce the original expressions of Erhenfest equations directly into the black hole phase transition research demonstrate again that black hole thermodynamics is closely related to the classical thermodynamics, which allows us to borrow techniques from classical thermodynamics to investigate the thermodynamics of black holes.

 \section*{Acknowledgements}
The authors would like to express their sincere gratitude to
the anonymous referees for their comments which help improve the
quality of the manuscript significantly. This research is supported by the National Natural Science
Foundation of China (Grant Nos.11235003, 11175019, 11178007). It is
also supported by \textquotedblleft Thousand Hundred
Ten\textquotedblright \,project of Guangdong Province and also by
Natural Science Foundation of Zhanjiang Normal University under
Grant No. QL1104.

\end{document}